\documentclass[11pt,english]{article}
\usepackage{graphicx}
\usepackage{xcolor}
\usepackage{amstext}
\usepackage{caption}
\usepackage{etoolbox}
\usepackage{makeidx}
\include{epsf}
\usepackage{amsfonts}
\usepackage{authblk} 
\usepackage{sectsty}
\usepackage{amsmath,amssymb,epsfig}
\usepackage{amscd}
\usepackage{amsthm}
\usepackage{mathrsfs}
\usepackage{dsfont}
\usepackage[applemac]{inputenc}
\usepackage[english]{babel}
\usepackage{enumitem} 
\usepackage[]{latexsym}
\usepackage{caption}
\usepackage{subfigure}
\usepackage{hyperref}
\usepackage{blindtext}
\usepackage{verbatim}
\usepackage{cancel}
\usepackage{cases}
\usepackage[notext]{stix}

\hypersetup{
pdftitle={},%
pdfauthor={},%
pdfsubject={},%
pdfkeywords={},%
colorlinks=true,%
linkcolor=blue,%
citecolor=crimson,%
linktocpage=true,%
hyperfootnotes=true,%
pageanchor=true
}

\definecolor{crimson}{rgb}{0.7, 0.08, 0.24}

\newcommand*{\affaddr}[1]{#1}
\newcommand*{\affmark}[1][*]{\textsuperscript{#1}}

\newtheorem*{proof*}{Proof}

\newcommand{\be}{\begin{equation}}

\newcommand{\ee}{\end{equation}}
\def\beqa{\begin{eqnarray}}
\def\eeqa{\end{eqnarray}}
\def\bean{\begin{eqnarray*}}
\def\eean{\end{eqnarray*}}

\newcommand{\dd}{\mathrm{d}}

\renewenvironment{thebibliography}[1]
         {\section*{References}\frenchspacing\small
          \begin{list}{[\arabic{enumi}]}
         {\usecounter{enumi}\parsep=2pt\topsep 0pt
         \settowidth{\labelwidth}{[#1]}
         \leftmargin=\labelwidth\advance\leftmargin\labelsep
         \rightmargin=0pt\itemsep=1pt\sloppy}}{\end{list}}

 \numberwithin{equation}{section}

\input xy
\xyoption{all}

\usepackage[normalem]{ulem}

\title{\textbf{\textsf{$(b,v)$-type variables for black to white hole transitions in effective loop quantum gravity}}\vspace{0.35cm}}

\author{
\textsf{Norbert Bodendorfer\affmark[1]\footnote{\texttt{norbert.bodendorfer@physik.uni-r.de}}, Fabio M. Mele\affmark[1]\footnote{\texttt{fabio.mele@physik.uni-r.de}}, and Johannes M\"unch\affmark[1]\footnote{\texttt{johannes.muench@physik.uni-r.de}}}\\
\affaddr{\affmark[1]\textsf{Institute for Theoretical Physics, University of Regensburg,}}\\
\affaddr{\textsf{93040 Regensburg, Germany}}\vspace{-0.5cm}
}

\begin{document}

\maketitle

\begin{abstract}
\textsf{Quantum gravity effects in effective models of loop quantum gravity, such as loop quantum cosmology, are encoded in the choice of so-called polymerisation schemes. Physical viability of the models, such as an onset of quantum effects at curvature scales near the Planck curvature, severely restrict the possible choices. An alternative point of view on the choice of polymerisation scheme is to choose adapted variables so that the scheme is the simplest possible one, known as $\mu_0$-scheme in loop quantum cosmology. There, physically viable models with $\mu_0$-scheme polymerise the Hubble rate $b$ that is directly related to the Ricci scalar and the matter energy density on-shell. Consequently, the onset of quantum effects depends precisely on those parameters. In this letter, we construct similar variables for black to white hole transitions modelled using the description of the Schwarzschild interior as a Kantowski-Sachs cosmology. The resulting model uses the $\mu_0$-scheme and features sensible physics for a broad range of initial conditions (= choices of black and white hole masses) and favours symmetric transitions upon invoking additional qualitative arguments. The resulting Hamiltonian is very simple and at most quadratic in its arguments, allowing for a straight forward quantisation.}

\end{abstract}

\section{Introduction}

Loop quantum gravity (LQG) is an approach to quantum gravity that directly quantises classical gravitational theories, such as standard general relativity in $3+1$ dimensions. It exists in Hamiltonian form \cite{ThiemannModernCanonicalQuantum, GambiniAFirstCourse}, as a path integral \cite{RovelliBook2}, as well as in the group field theory language \cite{OritiGroupFieldTheoryAsTheSecond}. Equivalence between the different formulations has not been shown in full generality so far, although much progress has been made when considering symmetry reduced situations such as cosmology, where at least qualitative agreement is reached \cite{AlesciANewPerspective, AlesciQuantumReducedLoopFull, GielenHomogeneousCosmologiesAs, OritiEmergentFriedmannDynamics, BeetleDiffeomorphismInvariantCosmological, BojowaldSphericallySymmetricQuantum, BojowaldSphericallySymmetricQuantumGeometryHamiltonian, DaporCosmologicalEffectiveHamiltonian, BIII, BLSI, BVI, BZI,BaytasEquivalenceofModels}.

For experimental tests of the theory, understanding symmetry reduced sectors is often enough, as the high energy densities necessary to induce strong quantum gravity effects e.g. appear near cosmological or black hole singularities. Hence, understanding the theory in simplified settings where such singularities still occur classically is a well motivated line of study. In the cosmological context, this idea has spawned the vast field of loop quantum cosmology, see \cite{BojowaldAbsenceOfSingularity, AshtekarMathematicalStructureOf, AshtekarQuantumNatureOf} for seminal papers and \cite{AshtekarLoopQuantumCosmology, SinghLoopQuantumCosmologyABrief} for reviews. 

In this letter, we will focus on the simplest possible black hole singularities, those of Schwarzschild black holes. Studying them with techniques similar to those of loop quantum cosmology is possible as the Schwarzschild interior can be rewritten as a Kantowski-Sachs cosmological model with the Schwarzschild variable $r$ as a time-like coordinate. This idea was follow up on in several papers already, however one often encountered physically insensible results \cite{ModestoLoopquantumblack,ModestoSemiclassicalLoopQuantum,ModestoBlackHoleinterior,BoehmerLoopquantumdynamics,ChiouPhenomenologicalloopquantum,ChiouPhenomenologicaldynamicsof,JoeKantowski-Sachsspacetimein} or had to deviate from the effective Hamiltonians typically arising in loop quantum cosmology \cite{AshtekarQuantumTransfigurationOf, AshtekarQuantumExtensionOf, BodendorferANoteOnTheHamiltonian}. As we will discuss in this letter, these problems can be evaded by choosing adapted variables similar to the $(b,v)$-variables in loop quantum cosmology \cite{AshtekarRobustnessOfKey} along with the simplest possible polymerisation scheme, which additionally allows for a straight forward construction of the quantum theory. We will be rather brief with technicalities in this letter. Detailed computations will appear in a companion paper \cite{BodendorferMassAndHorizon}. 

This letter is organised as follows:\\
Section \ref{sec:LQC} provides some background material on loop quantum cosmology and explains why it is physically sensible to use $\mu_0$ polymerisation schemes along with $b,v$ variables. Section \ref{sec:Classical} reviews the classical description of the Schwarzschild black hole interior as a Kantowski-Sachs cosmological model. Our new variables are motivated and chosen in section \ref{sec:Variables} and the effective Hamiltonian is derived. Physical predictions of the model are summarised in section \ref{sec:Physics}. Finally, we conclude in section \ref{sec:Conclusion}.

\section{$(b,v)$-variables in LQC} \label{sec:LQC}

Loop quantum cosmology (LQC) has originally been constructed as a mini-superspace quantisation of cosmological models, using some key concepts from full LQG \cite{BojowaldAbsenceOfSingularity, AshtekarMathematicalStructureOf, AshtekarQuantumNatureOf}. While heuristic derivations such as \cite{AshtekarQuantumNatureOf} argue that this should be understood as the continuum limit of a discretised full quantum gravity theory, it turns out that loop quantum cosmology can be best understood {\it and exactly derived} as a one-vertex ( = lattice point) truncation of a full discrete quantum gravity theory \cite{BIII, BVI}. Taking a continuum limit in such a theory then leads to quantitative changes in the predictions, but qualitative similarities \cite{BWI}. Thus, as is often acknowledged for various reasons, LQC-type models should be taken with a grain of salt and used only qualitatively unless derived in a continuum limit from a full theory. 

Having this in mind, it is easy to understand why the early LQC models \cite{BojowaldAbsenceOfSingularity, AshtekarMathematicalStructureOf} gave physically insensible results for the onset of quantum effects. As one uses only a single lattice point, holonomies of the Ashtekar-Barbero connection $A$ are evaluated along straight lines $\gamma$, parametrised by the variable $c$ as \cite{AshtekarMathematicalStructureOf}
\be
	h_\gamma(A) = \mathcal P \exp \left(\int_\gamma A \right) = \cos\left(\frac{\mu c}{2}\right) + 2 \sin\left(\frac{\mu c}{2}\right) (\dot \gamma^{a} \, {}^0\omega_a^i) \tau^i \label{eq:Areduced} ,
\ee
that run through all of the universe, see \cite{BIII} for a detailed construction. Here, by ``all of the universe'', we mean either a closed loop in a spatially compact universe such as a three-torus, or from boundary to boundary of a fiducial cell in the non-compact case. We denote by ${}^0\omega_a^i$ the fiducial co-triad, $\dot \gamma^a$ the (constant) tangent to $\gamma$, and $\mu := 2 \mu_0$ a free parameter that is fixed once and for all in the derivation. As a consequence of approximating field strengths via holonomies of closed loops, one finds that the gravitational part of the Hamiltonian constraint is modified as 
\be
	c \mapsto \sin(\mu_0 c) / \mu_0 \label{eq:Sub}
\ee
While \cite{AshtekarMathematicalStructureOf, AshtekarQuantumNatureOf} offered a detailed derivation of this procedure from full LQG arguments using \eqref{eq:Areduced}, the substitution \eqref{eq:Sub} has usually been adopted as a direct ``effective'' mean to access the quantum theory. 

Irrespective of how one arrives at \eqref{eq:Sub}, it is immediately clear that corrections to classical general relativity are suppressed only as long as $\mu_0 c \ll 1$. In the homogeneous, isotropic, and spatially flat context, we have $c \propto a \cdot b$, where $a$ is the scale factor describing the physical spatial extend of the universe, and $b\propto \dot a / a$ is the Hubble rate. Furthermore, the Ricci scalar is simply given by $R \propto b^2$ and the matter energy density also satisfies $\rho_m \propto b^2$. It follows that by taking $a$ large enough, we can encounter corrections to classical general relativity at arbitrarily low curvatures and matter energy densities and thus arrive at insensible physics.

Using a more elaborate argument based on the area gap of full LQG, it was proposed in \cite{AshtekarQuantumNatureOf} that instead of fixing $\mu=\text{const.}$ once and for all, one should rather introduce a dynamical quantity $\bar{\mu} \propto 1/a$. As a consequence, quantum effects are suppressed as long as $b\ll1$, which is sensible as quantum effects now become dominant at the Planck curvature or Planck energy density. The argument of \cite{AshtekarQuantumNatureOf} can be understood to lead to this result as follows: one demands that the integrated curvature evaluated via a closed loop holonomy around a plaquette of area $1$ in Planck units is cut off at value $1$ in Planck units. The Planck unit curvature cutoff follows heuristically.

For constant $\mu$, a quantum theory can be constructed using the square integrable functions on U$(1)$. Taking $\bar \mu$ to be a dynamical quantity poses several technical challenges for implementing it in a quantum theory, i.e. beyond the classical ``effective'' theory obtained via \eqref{eq:Sub}. 
In the context of loop quantum cosmology, this problem could be solved by substituting U$(1)$ with the Bohr compactification of the real line, see the discussion in \cite{AshtekarMathematicalStructureOf}. However, no analogue of the Bohr compactification is known for non-Abelian groups such as SU$(2)$, which calls this procedure into question as a means to obtain sensible physics from full LQG. 

It was noted in \cite{AshtekarRobustnessOfKey} that it may be a better idea to instead consider $b = \bar{\mu} c / \mu_0$ as a fundamental variable and build the quantisation on the canonical pair $\{b, v\} \propto 1$, where $v = a^3$ is the physical volume. In fact, the $\bar{\mu}$-scheme follows from substituting 
\be
	b \mapsto \sin(\mu_0 b) / \mu_0 \label{eq:SubB} \text{,}
\ee
which avoids the technical problems of using non-constant $\mu$ in the quantum theory. This idea can also be incorporated in the full theory by parametrising the full phase space of general relativity by similar variables \cite{BVI}.

This observation motivates the main goal of this paper, which is to find similar variables to describe physically sensible black to white hole transitions in LQG using a $\mu_0$ scheme.

\section{Classical setup} \label{sec:Classical}

As the material covered in this section is already known, we will be rather brief and refer to our companion paper \cite{BodendorferMassAndHorizon} for details. 
The most general ansatz for a static spherically symmetric metric is given by \cite{CampigliaLoopquantizationof,VakiliClassicalpolymerizationof}

\begin{equation}\label{eq:metricansatz}
\dd s^2 = - \bar{a}(r) \dd t^2 + N(r) \dd r^2 + 2 \bar{B}(r) \dd r \dd t + \mathscr b(r)^2 \dd \Omega_2^2 \;,
\end{equation} 
where $\dd \Omega_2$ denotes the metric on the $r,t = const.$ round 2-sphere. In the Schwarzschild interior, $\bar{a}(r), N(r) < 0$, and the $t$-direction is consequently non-compact and spacelike. Hence, it is convenient to define the integrated quantities 
$$
\sqrt{a} = \int_{0}^{L_o} \sqrt{\bar{a}} \;\dd t = L_o\sqrt{\bar{a}} , \quad B = \int_{0}^{L_o} \bar{B} \;\dd t = L_o \bar{B}  , \quad n = Na + B^2\;,
$$
\noindent
where $L_o$ is the coordinate size of a fiducial cell and we further define $\mathscr L_o = \int_{0}^{L_o} \;\dd t \left.\sqrt{\bar{a}}\right|_{r=r_{\text{ref}}}$. 

In terms of spherically symmetric connection variables with the gauge choice $B=0$, the metric reads (see e.g. \cite{ModestoLoopquantumblack})
\begin{equation}\label{eq:metricansatzconn}
ds^2 = -N_T^2(T) \,\dd T^2 + \frac{p_b^2(T)}{L_o^2 |p_c(T)|}\, \dd x^2 + |p_c(T)| \,\dd \Omega_2^2 ,
\end{equation}
\noindent
where we identified 
\begin{equation}\label{eq:metricmap}
T = r,\quad x = t ,\quad   |p_c| = \mathscr b^2 ,\quad p_b^2 = -a \mathscr b^2,\quad N = - N^2_T \text{.}
\end{equation} 
Under a scaling $L_o \mapsto \alpha L_o$ of the fiducial cell, the variables transform as
\begin{equation}\label{eq:sclaingcb}
b \longmapsto b  , \quad c \longmapsto \alpha c  , \quad p_b \longmapsto \alpha p_b  , \quad p_c \longmapsto p_c \;.
\end{equation}

The Hamiltonian constraint of this system reads 
\begin{equation}\label{eq:classHamconn}
H = N_T \mathcal{H}  , \quad \mathcal{H} = - \frac{b}{2 G \gamma^2 \text{sign}(p_c) \sqrt{|p_c|}} \left( 2 c p_c + \left(b+ \frac{\gamma^2}{b} \right) p_b \right) \approx 0 \;,
\end{equation}
where $\gamma \in \mathbb R \backslash \{0\}$ is the Barbero-Immirzi parameter and the equations of motions can be obtained using the Poisson brackets 
\be
	\left\{b, p_b\right\} = G \gamma  , \quad \left\{c, p_c\right\} = 2 G \gamma \;.
\ee
The equation of motion can be solved as
\begin{align}
b(T) = \pm \gamma\sqrt{A e^{-T}-1}  ,& ~~\quad c(T) = c_o e^{-2T} \\
p_b(T) = -\frac{2 c p_c}{b+ \frac{\gamma^2}{b}} = \mp \frac{2c_o p_c^o}{\gamma} \sqrt{\frac{e^T}{A} \left(1- \frac{e^T}{A}\right)}  ,& ~~\quad p_c(T) = p_c^o e^{2T} \;,
\end{align}
where one integration constant was eliminated using the Hamiltonian constraint. Using the identification $A = e^{-T_0}$, we can furthermore set $A$ to one by shifting the $T$-coordinate. 

Using the gauge choice and two variable redefinitions 
\begin{equation}\label{eq:classcoordinatechange}
	N_T = \frac{\gamma \;\text{sign}(p_c) \sqrt{|p_c|}}{b}, ~~\quad \quad~~ \tau = \sqrt{|p_c^o|} e^T , \quad y = \frac{2 c_o \sqrt{|p_c^o|}}{\gamma}x \;,
\end{equation}
we arrive at
\begin{equation}\label{eq:metricclfinal}
\dd s^2 = -\frac{1}{\frac{\sqrt{|p_c^o|}}{\tau}-1} \dd \tau^2 + \left(\frac{\sqrt{|p_c^o|}}{\tau} -1\right) \dd y^2 + \tau^2 \dd \Omega_2^2 \;
\end{equation}
and can identify $\sqrt{|p_c^o|} = 2M = R_{hor}$, where $M$ is the black hole mass and $R_{hor}$ the horizon radius. 

The system can also be described using the two Dirac observables 
\begin{equation}\label{eq:DOclass}
	\mathcal{R}_{hor} = \sqrt{|p_c|} \left(\frac{b^2}{\gamma^2} + 1\right) \overset{\text{on-shell}}{=} R_{hor}  , \quad \quad \mathcal{D} = c p_c \text{.}
\end{equation}
Due to the scaling properties $\mathcal{R}_{hor} \longmapsto \mathcal{R}_{hor} , ~ \mathcal{D} \longmapsto \alpha \mathcal{D} $ under a change of fiducial cell, only $\mathcal{R}_{hor}$ is physical. This last observation turns out to change in the quantum theory, where an additional fiducial cell independent Dirac observable that corresponds to the white hole mass can be constructed using polymerisation parameters that also scale under a fiducial cell change.

\section{New variables} \label{sec:Variables}

To compare and contrast with a recent work by the authors \cite{BodendorferEffectiveQuantumExtended}, we introduce the canonical pairs $(v_1,P_1)$, $(v_2,P_2)$ as
 \begin{align}
 \left(p_b\right)^2 = -8 v_2  &, \quad\quad |p_c| = \left(24 v_1\right)^{\frac{2}{3}}\;, 
 \label{connvaraibles1}
 \\
 b = \text{sign}(p_b)\; \frac{\gamma}{4}\; \sqrt{-8 v_2} \;P_2  &, \quad\quad c = -\text{sign}(p_c)\; \frac{\gamma}{8}\; \left(24 v_1\right)^{\frac{1}{3}}\; P_1  \;,
 \label{connvaraibles2}
 \end{align}
 so that $\left\{v_i, v_j\right\} = 0, ~ \left\{P_i, P_j\right\} = 0  ,~  \left\{v_i, P_j\right\} = \delta_{ij}$, where we set $G=1$ from now on. It follows that
  \begin{equation}\label{eq:HamclassvP}
	 H_{cl}=\sqrt{n}\mathcal{H}_{cl} ,\quad \mathcal{H}_{cl} = 3v_1P_1P_2+v_2P_2^2-2 \approx 0\;,
 \end{equation}
 where $n = N a $ is a Lagrange multiplier. The metric components can be reconstructed as
  \begin{equation}
 a = \frac{v_2}{2} \left(\frac{2}{3 v_1}\right)^{\frac{2}{3}} ,\quad \mathscr b = \left(\frac{3 v_1}{2}\right)^{\frac{1}{3}} \;.
 \end{equation}
 
It was observed in \cite{BodendorferEffectiveQuantumExtended} that a $\mu_0$-scheme polymerisation of these variables leads to a maximal value of the Kretschmann scalar depending on the chosen initial conditions, which is physically undesirable. This can be remedied as follows. 

In the variables $(v_1,P_1,v_2,P_2)$, the on-shell expression for the Kretschmann scalar reads
\be\label{KretschmannvP}
\mathcal K(v_1,P_1,v_2,P_2)=12\left(\frac{3}{2}v_1\right)^{\frac{2}{3}}P_1^2P_2^2\;.
\ee
This suggests to use a power of $\mathcal K(v_1,P_1,v_2,P_2)$ as a canonical variable, similar to the Ricci scalar $R \propto b^2$ in LQC. To this end, we introduce the new canonical variables 
\be\label{vPtokj}
v_k=\left(\frac{3}{2}v_1\right)^{\frac{2}{3}}\frac{1}{P_2}  ,\quad\quad  v_j=v_2-\frac{3v_1P_1}{2P_2} ,\quad\quad k=\left(\frac{3}{2}v_1\right)^{\frac{1}{3}}P_1P_2 ,\quad\quad j=P_2 \;.
\ee
with non-vanishing Poisson brackets
\be
\begin{aligned}
	&\{v_k,k\}=1, \quad \quad\{v_j,j\}=1.
\end{aligned}
\ee
The main reason for this variable choice is the observation that $\mathcal K \propto k^2$, i.e. $\mu_0$ scheme polymerisations of $k$ are expected to lead to an upper bound for $\mathcal K$ determined by the choice of $\mu_0$. In Planck units, the natural choice $\mu_0 \approx 1$ would lead to an upper bound given by the Planck curvature. 

Following standard procedures, we derive an effective quantum theory via substituting\footnote{Let us remark that many alternative proposals of polymerisation have been considered in the literature. These include choosing different functions or polymerising only parts of the phase space or different choices for the polymerisation scales (see e.g. \cite{ModestoSemiclassicalLoopQuantum,BojowaldCovarianceInModels,BojowaldEffectiveLineElements,BenAchourPolymerSchwarzschildBlack, AssanioussiPerspectivesOnThe} and references therein). Such different models can be motivated by physical inputs or full theory based results and arguments like general covariance and anomaly-free realisations of the constraint algebra. In particular, the analysis of the issue of gauge anomalies in the spherically symmetric setting as well as its implication in a static framework has led to new models of LQG black hole  \cite{BojowaldEffectiveLineElements,BenAchourPolymerSchwarzschildBlack}. Here, for simplicity, we do not consider such alternative choices and rather focus on the simplest choice of sin-polymerisation which, for the variables introduced in the present model, results into a physically reasonable effective Schwarzschild spacetime discussed in the next section. We will briefly comment on the extension to generic $t$- and $r$-dependent spherical symmetry in the conclusion section, while leaving its deserved in depth study and the analysis of the resulting constraint algebra for future work.} 
\begin{align}
&k\longmapsto \frac{\sin(\lambda_k\,k)}{\lambda_k} , \quad \quad j\longmapsto \frac{\sin(\lambda_j\,j)}{\lambda_j}\;, \label{polymerisation}
\end{align}
\noindent
where we keep $\lambda_j$ and $\lambda_k$ constant (corresponding to two independent choices for $\mu_0$ in the two variable sectors). Since the arguments of the $\sin$ functions should not scale under fiducial cell rescalings, we need to impose 
\be\label{fidcellpolscale}
	\lambda_k\longmapsto\lambda_k,\qquad \lambda_j\longmapsto \alpha\,\lambda_j
\ee
under fiducial cell rescalings. Consequently, $\lambda_j$ can enter physical results only in ratios with other similarly scaling quantities.

From the purely classical model, the on-shell expression for $k$ and $j$ reads
\be\label{classicalonshell}
k(\mathscr b)=\left(\frac{D}{\mathscr L_o}\right)^{\frac{3}{2}}\frac{C}{\mathscr b^3}=\frac{2M_{BH}}{\mathscr b^3},\quad\quad \mathscr L_o\,j(\mathscr b)=\left(\frac{D}{\mathscr L_o}\right)^{\frac{1}{2}}\frac{1}{\mathscr b} \text{,}
\ee
where $\mathscr L_0 = \sqrt{n} = \text{const}$ has been adopted and $C$, $D$ are the two integration constants (see \cite{BodendorferMassAndHorizon} for details). It follows that, up to the $D$-dependence discussed later, the scale $\lambda_j$ controls quantum corrections for small radii of the two-spheres.

The polymerised effective Hamiltonian then reads 

\be\label{eq:hampolyjk}
H_{\text{eff}}=\sqrt{n}\,\mathcal H_{\text{eff}},\quad\quad\mathcal H_{\text{eff}}=3v_k\frac{\sin(\lambda_k\,k)}{\lambda_k}\frac{\sin(\lambda_j\,j)}{\lambda_j}+v_j\frac{\sin^2(\lambda_j\,j)}{\lambda_j^2}-2\approx0\;,
\ee

\noindent
and the corresponding equations of motion are given by

\be \label{eq:EOM}
\begin{cases}
	v_k'=3\sqrt{n}\,v_k\,\cos(\lambda_kk)\,\frac{\sin(\lambda_jj)}{\lambda_j}\\
	v_j'=3\sqrt{n}\,v_k\,\frac{\sin(\lambda_kk)}{\lambda_k}\,\cos(\lambda_jj)+2v_j\,\sqrt{n}\,\frac{\sin(\lambda_jj)}{\lambda_j}\,\cos(\lambda_jj)\\
	k'=-3\sqrt{n}\,\frac{\sin(\lambda_kk)}{\lambda_k}\,\frac{\sin(\lambda_jj)}{\lambda_j}\\
	j'=-\sqrt{n}\,\frac{\sin^2(\lambda_jj)}{\lambda_j^2}\\
\end{cases}\;.
\ee

As discussed in section \ref{sec:LQC}, choices of variables before polymerisation can be translated in the choice of a $\bar{\mu}$-scheme. In case of the variables advocated in this section, the corresponding scheme turns out to be rather complicated as $\bar{\mu}$ would depend on both connection and triad variables \cite{BodendorferMassAndHorizon}.  

\section{Physical predictions} \label{sec:Physics}

\subsection{Spacetime}

The equations of motion \eqref{eq:EOM} can be solved as

\begin{align}
v_k(r)&=\frac{2DC^2\lambda_k^2\sqrt{n}^3}{\lambda_j^3}\frac{\frac{\lambda_j^6}{16C^2\lambda_k^2n^3}\left(\frac{\sqrt{n}\,r}{\lambda_j}+\sqrt{1+\frac{nr^2}{\lambda_j^2}}\right)^6+1}{\left(\frac{\sqrt{n}\,r}{\lambda_j}+\sqrt{1+\frac{nr^2}{\lambda_j^2}}\right)^3}\;,\label{effdynsol1}\\
v_j(r)&=2n\left(\frac{\lambda_j}{\sqrt{n}}\right)^2\left(1+\frac{nr^2}{\lambda_j^2}\right)\left(1-\frac{3CD}{2\lambda_j}\frac{1}{\sqrt{1+\frac{nr^2}{\lambda_j^2}}}\right)\;,\label{effdynsol2}\\
k(r)&=\frac{2}{\lambda_k}\cot^{-1}\left(\frac{\lambda_j^3}{4C\lambda_k\sqrt{n}^3}\left(\frac{\sqrt{n}\,r}{\lambda_j}+\sqrt{1+\frac{nr^2}{\lambda_j^2}}\right)^3\right)\;,\label{effdynsol3}\\
j(r)&=\frac{1}{\lambda_j}\cot^{-1}\left(\frac{\sqrt{n}r}{\lambda_j}\right)+\frac{\pi}{\lambda_j}\theta\left(-\frac{\sqrt{n}r}{\lambda_j}\right)\;,\label{effdynsol4}
\end{align}
as rewritten in terms of $a$ and $\mathscr b$ as function of $x := \mathscr L_o r/\lambda_j$ as
\begin{align}
\mathscr b^2(x)&=\frac{1}{2} \left(\frac{\lambda_k}{M_{BH} M_{WH}}\right)^{\frac{2}{3}}\frac{1}{\sqrt{1+x^2}}\frac{M_{BH}^2\left(x+\sqrt{1+x^2}\right)^6+M_{WH}^2}{\left(x+\sqrt{1+x^2}\right)^3}\;,\\
\frac{a(x)}{\lambda_j^2}&=2 \left(\frac{M_{BH} M_{WH}}{\lambda_k}\right)^{\frac{2}{3}}\left(1-\left(\frac{M_{BH} M_{WH}}{\lambda_k}\right)^{\frac{1}{3}}\frac{1}{\sqrt{1+x^2}}\right)\frac{\left(1+x^2\right)^{\frac{3}{2}}\left(x+\sqrt{1+x^2}\right)^3}{M_{BH}^2\left(x+\sqrt{1+x^2}\right)^6+M_{WH}^2}\,.
\end{align}
Again, we obtain two integration constants $C$ and $D$. They are reflected in the values of the two fiducial cell independent Dirac observables 
\begin{align}
2\mathcal{M}_{BH}&=\frac{\sin(\lambda_kk)}{\lambda_k}\cos\left(\frac{\lambda_kk}{2}\right)\left(\frac{2v_k}{\lambda_j\cot\left(\frac{\lambda_jj}{2}\right)}\right)^{\frac{3}{2}}\;,\\
2\mathcal{M}_{WH}&=\frac{\sin(\lambda_kk)}{\lambda_k}\sin\left(\frac{\lambda_kk}{2}\right)\left(\frac{2v_k}{\lambda_j}\cot\left(\frac{\lambda_jj}{2}\right)\right)^{\frac{3}{2}}\;,
\end{align} 
as
\be\label{eq:CDjk}
C=\frac{\lambda_j^3}{4\lambda_k\sqrt{n}^3}\frac{M_{WH}}{M_{BH}},\qquad D=\sqrt{n}\left(\frac{8 \lambda_k\sqrt{n}^3}{\lambda_j^3}\frac{M_{BH}^2}{M_{WH}}\right)^{\frac{2}{3}}\;. 
\ee

\begin{figure}[t!]
	\centering\includegraphics[scale=0.5]{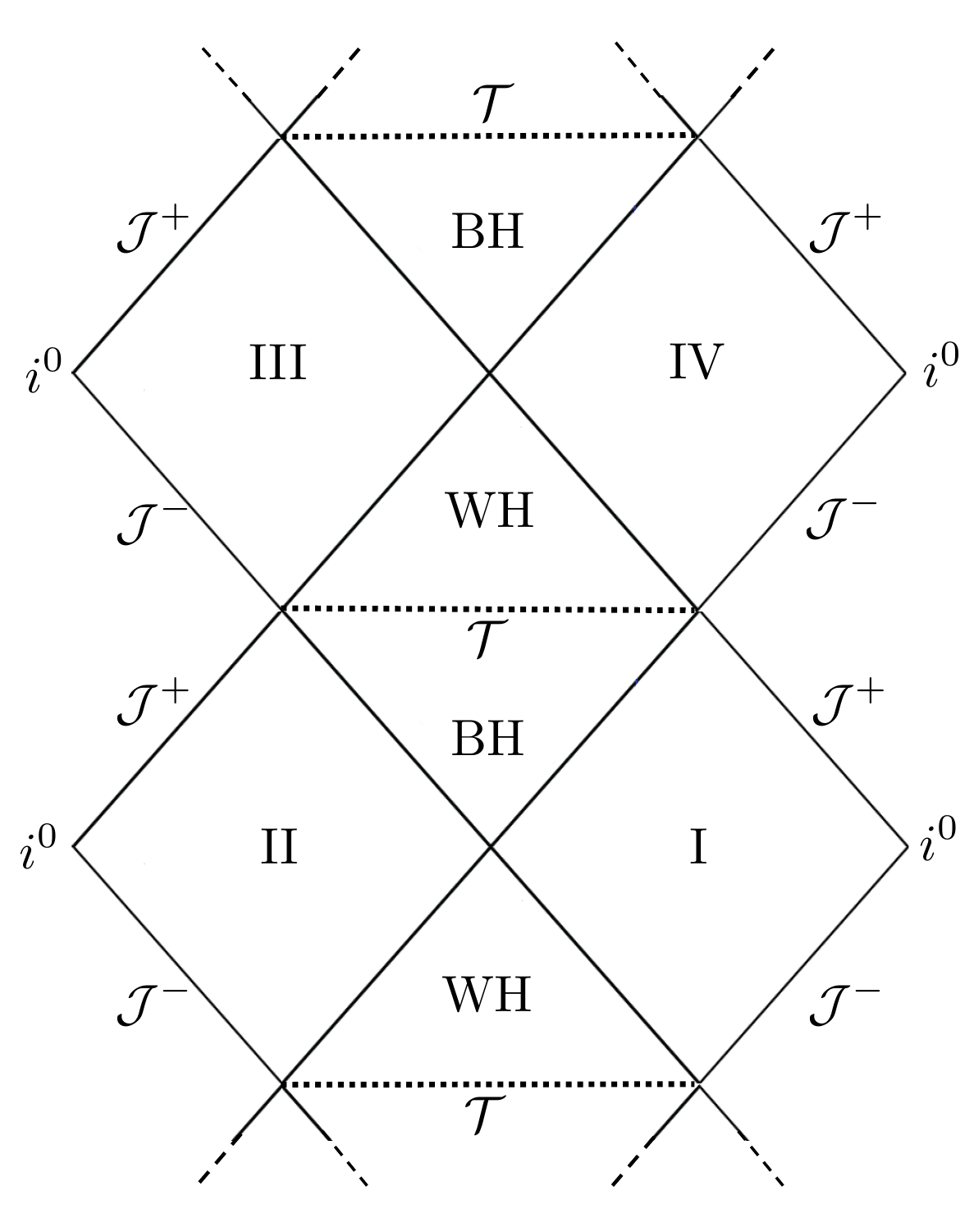}
	\caption{Penrose diagram for the Kruskal extension of the full quantum corrected polymer Schwarzschild spacetime.}
	\label{Penrosediag2}
\end{figure}

A detailed construction of the metric in the far future and far past after the black to white hole transition leads to the Penrose diagram in Fig. \ref{Penrosediag2}.
The key observations entering its construction are as follows (see \cite{BodendorferMassAndHorizon} for details).
a) Both in the asymptotic future and past, the spacetime is asymptotically flat. b) Far away from the transition surface connecting the black and while hole regions, the metric is approximately classical and corresponds to Schwarzschild spacetimes with masses $M_{BH}$ and $M_{WH}$, which can be chosen arbitrarily as initial conditions. 
c) Following the spacetime evolution from the black hole classical regime to the white hole classical regime up to the same value of $\mathscr b$ leads to 
\begin{equation}
v_j\mapsto v_j \quad ,\quad v_k \mapsto v_k\quad, \quad k \mapsto \frac{\pi}{\lambda_k}- k \quad ,\quad  j \mapsto \frac{\pi}{\lambda_j} - j \;.
\end{equation}
The Dirac observables transform under this process as
\begin{equation}
	 \mathcal{M}_{BH} \longmapsto \mathcal{M}_{WH} , \quad  \mathcal{M}_{WH} \longmapsto  \mathcal{M}_{BH} \;.
\end{equation}
leading to an infinite oscillation between Schwarzschild spacetimes with masses $M_{BH}$ and $M_{WH}$.

\subsection{Onset of quantum effects}

As expected, we observed numerically that for a very broad range of $M_{BH}$ and $M_{WH}$ (with numerically stable results for $M_{BH},M_{WH} < 10^{20}$), the maximum value of the Kretschmann scalar is bounded by approximately the Planck curvature for $\lambda_k \approx 1$, see \cite{BodendorferMassAndHorizon} for details. 

As for the onset of quantum effects, it is clear from the polymerisation \eqref{polymerisation} that the spacetime is approximately classical as long as 
\be
\frac{\mathscr L_o\,r}{\lambda_j}\gg1,\qquad\frac{2r^3}{C\lambda_k}\gg1. \label{eq:Conditions}
\ee
The second condition can be rewritten as 
\begin{equation}\label{eq:curvscaleBH}
\mathcal{K}_{cl}^{BH} = \frac{48 M_{BH}^2}{\mathscr b_+^6} \ll \frac{48}{\lambda_k^2} \;,
\end{equation}
where $\mathscr b_+ = \mathscr b(r\rightarrow +\infty)$ is the value of $\mathscr b$ in the black hole region, but far away from the transition surface, and corresponds to an onset of quantum effects when the Kretschmann scalar becomes close to the scale $1/\lambda_k^2$. On the white hole side, this equation becomes

\begin{equation}\label{eq:curvscaleWH}
\mathcal{K}_{cl}^{WH} = \frac{48 M_{WH}^2}{\mathscr b_-^6} \ll \frac{48}{\lambda_k^2} \;,
\end{equation}
with $\mathscr b_-= \mathscr b(r\rightarrow -\infty)$ denoting the corresponding value in the white hole region. 

For the first condition in \eqref{eq:Conditions} corresponding to small radius corrections, it can be shown that their onset is always after large curvature effects originating from the second condition for the range 
\be\label{eq:massratio}
	\frac{1}{8}<\frac{M_{WH}}{M_{BH}} < 8 \;
\ee
of initial conditions. Outside of this range, one encounters an onset of quantum effects at curvatures much lower than the Planck curvature. This can be understood from rewriting the first condition of \eqref{eq:Conditions} as (and equivalent on the white hole side)
\be
\frac{2 M_{BH}}{\mathscr b_+^3} \ll \frac{1}{4 \lambda_k} \frac{M_{WH}}{M_{BH}} ,\quad\quad \frac{2 M_{WH}}{\mathscr b_-^3} \ll \frac{1}{4 \lambda_k} \frac{M_{BH}}{M_{WH}}, \label{eq:DinPolymer}
\ee
which is a curvature scale depending on the mass ratio $M_{WH}/M_{BH}$. 

Although the upper curvature bound is fine for all mass ratios, the natural conclusion of the above observation about the onset of quantum effects is that physically reasonable black to white hole transitions preferred by the model are those where the masses do not change significantly. Rather, choosing $M_{WH} = M_{BH}$ perfectly aligns both types of corrections, making them both appear at high curvatures. From a physical point of view, one may expect that no mass is gained or lost in a black to white hole transition, showing that such a restriction of the initial conditions may be sensible. Since the quantum theory corresponding to the exceptionally simple Hamiltonian \eqref{eq:hampolyjk} can be explicitly constructed using standard LQC methods (see e.g. \cite{BodendorferEffectiveQuantumExtended}), one may also address this question using wave packets. Due to $\mathcal M_{WH}$ and $\mathcal M_{BH}$ not Poisson-commuting, one can not to specify both of them simultaneously with arbitrary precision, which may affect the discussion. 

We have not been able to avoid the $D$-dependence in \eqref{eq:DinPolymer} by another choice of variables while keeping \eqref{eq:curvscaleBH} and \eqref{eq:curvscaleWH} as is. 
Making $\lambda_j$ $D$-dependent as a choice of polymerisation scheme, following the ideas of \cite{AshtekarQuantumTransfigurationOf, AshtekarQuantumExtensionOf}, is problematic for various reasons, and changes the equations of motion \cite{BodendorferANoteOnTheHamiltonian}, so that no immediate conclusions can be drawn\footnote{Further drawbacks originating from the polymmerisation strategy adopted in \cite{AshtekarQuantumTransfigurationOf, AshtekarQuantumExtensionOf} have been pointed out in the literature. These concern issues with general covariance \cite{BojowaldComment2AOS} or departures from the expected asymptotic Minkowski structure \cite{BrahmaCommentonAOS,FaraoniUnsettlingThePhysics} (on the latter point see also \cite{AshtekarPropertiesOfRecent} for completeness).}. 
It is however possible to obtain sensible small 2-sphere radius corrections by restricting the initial conditions to $D\approx1$, which leads to $M_\text{WH} \propto M_\text{BH}^2$ \cite{BodendorferMassAndHorizon}. Nevertheless, the symmetric bounce above, where the two types of corrections reduce to large curvature corrections, seems more natural to us.

\section{Conclusion} \label{sec:Conclusion}

We have presented a new model for black to white hole transitions inspired by LQG. The physical idea entering our model is to construct sensible quantum corrections appearing once the spacetime curvature becomes close to the Planck curvature by polymerising adapted variables. Our model satisfies all criteria of physical viability (sensible onset of quantum effects, Planckian upper bounds on curvature scalars, possibility of symmetric bounce), as e.g. spelled out in \cite{AshtekarQuantumTransfigurationOf}. It does so by using a simple $\mu_0$-scheme, i.e. constant polymerisation scales that can be immediately transferred to a quantum theory. To the best of our knowledge, the presented model is currently the only one in the literature satisfying all of the above. 

For future work, it would be interesting to study the quantum theory obtained from \eqref{eq:hampolyjk} and embed it into full quantum gravity via the methods of \cite{BIII, BVI}. Once this is done, it is possible to study coarse graining following \cite{BodendorferCoarseGrainingAs, BWI}, which may affect some of the physical predictions. After all, the model discussed here is expected (by analogy with \cite{BIII, BVI}) to correspond to a one-vertex truncation of a full quantum gravity theory, thus neglecting possible effects of the continuum limit as illustrated in \cite{BWI}.

Moreover, as already mentioned in Sec. \ref{sec:Variables}, it would be interesting to extend the variables presented here beyond the static Schwarzschild case by considering a generic spherically symmetric $t$- and $r$-dependent line element as starting point. Due to their geometric interpretation as being respectively related to the Kretschmann scalar and the angular components of the extrinsic curvature, the momenta $k$ and $j$ can in principle be straightforwardly computed also in the generic spherically symmetric case. Less straightforward would be the construction of the corresponding conjugate variables but, modulo computational difficulties, still possible for instance via generating function methods. A successful construction would not only be a necessary step towards more realistic models of quantum corrected black holes but would allow us to study also key questions about the now non-trivial algebra of constraints and general covariance along the lines of \cite{BojowaldEffectiveLineElements,BenAchourPolymerSchwarzschildBlack}. In this respect, let us mention that the considerations in \cite{BojowaldEffectiveLineElements,BenAchourPolymerSchwarzschildBlack} are based on connection variables and mainly limited to their $\mu_o$-scheme (see however \cite{BenAchourCovarianceSelfDual} for a proposal of $\bar\mu$-scheme in polymer black holes using self-dual variables). A better understanding of the relation of our new variables, eventually extended to the $t$- and $r$-dependent case, with connection variable-based polymerisation schemes might be therefore desirable for comparison. The fact that the polymerisation strategy in \cite{BojowaldEffectiveLineElements,BenAchourPolymerSchwarzschildBlack} also turns out to focus on the angular component of the extrinsic curvature together with the interpretation of our $k$ momentum in terms of spacetime scalars is in a sense encouraging. A detailed study of the class of polymerisation fucntions (not necessarily of the sin form) compatible with anomaly-free considerations in our framework as well as their consequences for the effective spacetime structure are left for future investigations.

\section*{Acknowledgements}

The authors were supported by an International Junior Research Group grant of the Elite Network of Bavaria.

\end{document}